\documentclass[12pt,preprint]{aastex}

\shorttitle{Image of Fomalhaut Dust Ring}
\shortauthors{Marsh et al.}

\begin{document}

\title{Image of Fomalhaut Dust Ring at 350 $\mu$m: Relative Column 
    Density Map Shows Pericenter-Apocenter Asymmetry}

\author{K. A. Marsh\altaffilmark{1,2}, T. Velusamy\altaffilmark{1,3}, 
C. D. Dowell\altaffilmark{1,4}, K. Grogan\altaffilmark{1,5}, and
C. A. Beichman,\altaffilmark{6}}

\altaffiltext{1}{Jet Propulsion Laboratory, 4800 Oak Grove Drive, Pasadena,
CA 91109}
\altaffiltext{2}{E-mail: Kenneth.A.Marsh@jpl.nasa.gov}
\altaffiltext{3}{E-mail: Thangasamy.Velusamy@jpl.nasa.gov}
\altaffiltext{4}{E-mail: cdd@submm.caltech.edu}
\altaffiltext{5}{E-mail: Keith.Grogan@jpl.nasa.gov}
\altaffiltext{6}{California Institute of Technology 100-22, Pasadena, CA 91125;
E-mail: chas@ipac.caltech.edu}

\pagebreak

\begin{abstract}
We have imaged the circumstellar disk of Fomalhaut at 350 
$\mu$m wavelength, using  SHARC II at the Caltech Submillimeter Observatory.
The spatial resolution of the raw images ($9''$) 
has been enhanced by a factor of three using the 
{\tt HiRes} deconvolution procedure. We find that at this 
wavelength and signal to noise ratio ($\sim 12$), the 
observed morphology is that of a simple inclined ring ($i \simeq 70^\circ$), 
with little or no other apparent structure---this is the first observation that
shows clearly the ring morphology of the disk.
We have combined our 350 $\mu$m data with {\em Spitzer Space Telescope\/} 
images at 24, 70, and 160 $\mu$m in order to estimate the 
2-dimensional spatial variation of relative column density 
(``tau map") using our 
{\tt DISKFIT} procedure. The tau 
map is based on the following physical assumptions: (1) 
the wavelength variation of opacity is the same throughout 
the disk, (2) the radial variation of dust temperature is 
dictated by the energy balance of individual grains in the 
stellar radiation field, and (3) the vertical scale height 
of the disk follows a power-law radial variation. The results 
confirm the ring-like morphology, but also show that the geometric center
is displaced from the star by about 8 AU and that the ring has an 
apocentric enhancement of approximately 14\% in integrated column density.
If we interpret the displacement 
in terms of elliptical orbital motion due to gravitational perturbation by 
an unseen planet, then the implied forced eccentricity is $\sim0.06$; dynamical
modeling then predicts an apocentric density enhancement consistent
with that inferred from the tau map.
\end{abstract}

\keywords{circumstellar matter --- planetary systems --- stars: individual
(\object{Fomalhaut})}

\section{Introduction}

The study of dusty debris disks around main sequence stars
can provide clues to the possible presence of orbiting bodies
such as planets, the gravitational effects of which have 
predictable effects on the orbital dynamics of dust particles (see, for
example, \citet{qui02,wil02}). One system of considerable interest is
Fomalhaut ($\alpha$ PsA), of which submillimeter-wave mapping
at 450 and 850 $\mu$m has
shown the presence of structure dominated by a pair of intensity maxima
interpreted as the ansae of an inclined ring 
\citep{hol03}.  Observations at 24, 70, and 160 $\mu$m with 
{\em Spitzer\/} \citep{wer04} show that the relative intensities of the ansae
are wavelength-dependent \citep{sta04}.

The spatial variation of dust column density in the disks provides information
about the perturbing bodies when interpreted using dynamical models.
However, since the observed (intensity) images are sensitive to variations in 
dust temperature as well as column density,
we clearly need a tool to separate these effects. With this in mind,
we have developed an inversion procedure ({\tt DISKFIT}) designed to estimate
the 2-dimensional spatial distribution of optical depth (or relative column 
density) of a circumstellar disk in either the sky plane or disk plane, 
given a set of observed images
at multiple wavelengths, the corresponding point source response functions
(PSFs), and certain model assumptions.  We have applied this procedure
to our new 350 $\mu$m image of Fomalhaut in conjunction with {\em Spitzer\/}
data.

\section{Observations and Data Reduction}

Fomalhaut was observed at 350 $\mu$m with the SHARC II
camera \citep{dow03} at the Nasmyth focus of the Caltech Submillimeter
Observatory (CSO) on 2004 Sep 18 UT.  Of a total period of 4.7 hr on source, 
three hour-long integrations between 8.5 and 12.3 h UT provided the
best signal-to-noise due to low atmospheric opacity (0.039 at 225 GHz,
1.05 at 350 $\mu$m; both at zenith), remarkable atmospheric stability, and
relatively high source elevation ($27^\circ$ to $41^\circ$).  Our final map
therefore consists of a coaddition of the images from that period,
with a total integration time of 3.0 hr.

The telescope executed a Lissajous scanning pattern
with peak-to-peak amplitude $60''$ and period 20 s in the azimuth direction,
and peak-to-peak amplitude $30''$ and period 14.14 s in the elevation
direction.  Of the 12 rows of the detector array, only rows 1--8 were
operational for this run; at a plate scale of 4\farcs57
${\rm pixel}^{-1}$, the instantaneous field of view was then 
$146'' \times 37''$.
The bolometer signals were sampled at 28 Hz and then reduced with the
{\tt CRUSH} software program (A. Kovacs, in preparation \footnote
{see http://www.submm.caltech.edu/$\sim$sharc/crush/index.htm}).
We used the ``-deep" option which takes a more
aggressive approach at modeling atmospheric and instrumental effects, but
found the results to be relatively insensitive to this choice,
perhaps due to the atmospheric
stability.  Automatic correction was made
for the atmospheric opacity 
and increased gain of the detectors under reduced atmospheric
loading, the latter being a $\sim6$\% effect implemented using a lookup table. 

Absolute calibration
was accomplished with hourly, interspersed observations of
Uranus and
Neptune.  Uranus (assumed flux 237 Jy ${\rm beam}^{-1}$, 
250 Jy total)
was observed at elevations of $59^\circ$ and $31^\circ$, and Neptune
(assumed flux 92 Jy ${\rm beam}^{-1}$, 94 Jy total) was observed at 
elevations of $42^\circ$ and $29^\circ$.  The resultant four scale
factors from instrument units to Jy ${\rm beam}^{-1}$ had a relative RMS of 
7\%.  We estimate an absolute flux calibration accuracy of 20\% and
pointing accuracy of $2''$.
The PSF (FWHM $\simeq9''$) was obtained from observations of 
Neptune, appropriately rotated and coadded so as to recreate accurately the 
rotational smearing due to the changing parallactic angle during the
Fomalhaut observations.  

The upper panel of Figure \ref{fig1} shows the resulting image of Fomalhaut at 
350 $\mu$m.  The integrated flux density is $1.18\pm0.24$ Jy.  In order 
to enhance the spatial resolution of the image,
we have deconvolved it using the {\tt HiRes} implementation \citep{bac05} of the
Maximum Correlation Method \citep{aum90}.  The result is
shown in the lower panel of Figure \ref{fig1}, the spatial resolution of which 
is approximately $3''$.  It is apparent that
the source is strikingly symmetrical, suggestive of a simple ring.  This
is, in fact, the first observation which clearly shows the ring-like
morphology.  A key question, though, is whether the dust ring is truly
uniform or has density variations. We have addressed this issue by
combining our 350 $\mu$m image with other data using our {\tt DISKFIT}
inversion procedure. 

\section{Estimation of Relative Column Density Distribution}

The principal model assumptions involved in {\tt DISKFIT} are:

\begin{enumerate}
\item  The wavelength variation of opacity, $\kappa(\lambda)$, is the same 
throughout the disk;  we have assumed a power-law
variation of the form $\kappa(\lambda) \propto\lambda^{-\beta}$.

\item The radial variation of dust temperature is given by the
energy balance of individual grains \citep{bac93},
based on a grain size-scale parameter, $\lambda_0$, which 
represents the wavelength above which the grains radiate inefficiently.

\item  The vertical scale height of the disk follows a power-law variation
with respect to radial distance, $r$.  In terms of the opening
angle, $\psi$, corresponding to 1 scale height above and below midplane, 
we have assumed $\psi(r) = \psi_{100} (r_{[{\rm AU}]}/100)^\gamma$, where 
$\psi_{100}$ is the opening angle at $r=100$ AU.
\end{enumerate}

\noindent There are two main steps in the inversion procedure:
\smallskip

\noindent {\bf Step 1: Parameter Estimation.} In this initial step, we
obtain estimates of some preliminary
parameters necessary for the optical depth mapping.  These parameters
consist of $\beta$, $\psi_{100}$, and $\gamma$ (defined above), 
the disk inclination, $i$, with respect to the sky plane, and the 
position angle, $\theta$,
of the tilt axis.  We base these estimates on a simple model involving
a circular disk of inner and outer radii
$r_{\rm in}$ and $r_{\rm out}$, respectively, with a power-law radial
dependence of optical depth (normal to disk plane), 
represented by $\tau(r) \propto r^{-\alpha}$.
Allowance is made for the possibility that the center of
the dust ring may be displaced from the star, as an approximate way of
taking into account the effects of orbital ellipticity. 
A maximum likelihood estimate is then made
of the full set of parameters characterizing this model, using the
set of observed images and PSFs, taking account of the measurement noise
in each observed image, and making allowance for the presence of small
errors in the position calibration of each observed image.

\noindent {\bf Step 2: Inversion for Optical Depth Map.}
Based on the estimated values of $i, \theta, \psi_{100}, \gamma$, and $\beta$
from the previous step, the
full inversion for the 2-dimensional distribution of line-of-sight optical depth
(referred to as a ``tau map") in the
plane of the sky is accomplished using an algorithm which is similar
mathematically to the Richardson-Lucy procedure \citep{ric72,luc74}.
Since the PSFs at each wavelength are
implicitly deconvolved using the prior knowledge of positivity
of optical depth, some superresolution is obtained.  The output is a map
of the line-of-sight optical depth at an arbitrary reference wavelength
(equivalent to relative column density) and the associated uncertainty map.
As a final step, we can transform this sky projection into a pole-on view
using a linear estimation procedure which takes into account the flared
geometry.

\subsection{Application to Fomalhaut}

The input data for the parameter-fit step consisted of
our 350 $\mu$m image and {\em Spitzer\/} images at 160 and 70 $\mu$m from the
Multiband Imaging Photometer for {\em Spitzer\/} (MIPS)---see \citet{rie04}. 
Although a MIPS image at 24 $\mu$m was also available, we did not include it
in this step since it contains a significant response to
the presence of an additional warm dust component \citep{sta04} which was not
included in the simple ring model. Both large (blackbody) and intermediate-sized
($\lambda_0=27$ $\mu$m; \citet{bac93}) grains were considered for the purpose
of the dust temperature calculation.  Although the
results were not strongly dependent on the assumed grain properties, the best 
fit was obtained under the blackbody assumption, consistent with the presence
of large grains as inferred by \citet{chi90}.

The complete set of estimated parameter values is given in Table \ref{table1},
with error bars representing the effect of random measurement noise;
$\Delta x$ and $\Delta y$ represent the estimated offsets of the center
of the dust ring with respect to the star, in a right-handed coordinate 
system in which the $y$-axis corresponds to the disk tilt axis (positive = 
NNW). We have verified that the results
are unaffected by the possible effects of imperfect removal of an instrumental
near-IR leak in the 160 $\mu$m image by repeating the fit excluding the 
160 $\mu$m data; there was no significant difference in the estimated 
parameters.

Figure \ref{fig2} shows the residual images from the 
uniform-ring model fit as compared to the observed images; each plot represents
a slice along the major axis of the tilted ring.  The
residuals are more or less consistent with the estimated background noise
and correspond to a $\chi_\nu^2=1.56$.  The fact that the latter value is
somewhat larger than unity suggests the possibility of a slight deviation
from a uniform ring.

The next step (tau map estimation) involved reprocessing the raw images, making
use of the above estimates of $i, \theta, \psi_{100}, \gamma$, and $\beta$.
The input data were the observed images at all four available wavelengths
(350, 160, 70, and 24 $\mu$m); the resulting tau map in the
sky-plane projection is shown in the left hand panel of
Figure \ref{fig3}. In this figure, the line-of-sight optical depth variation
is expressed at a reference wavelength of 24 $\mu$m; the spatial resolution
is approximately 22 AU.  The peak values of optical depth in the ring
are $(6.61\pm0.28)\times 10^{-3}$ and $(6.56\pm0.33)\times 10^{-3}$ at the
SSE and NNW ansae, respectively.
The density-weighted dust temperature in the ring
is 42 K. Since the uncertainty
of the estimated optical depth increases with radial distance,
the map has been truncated at 310 AU, at which the
uncertainty is twice that at ring maximum.

It is apparent that the SSE peak on the tau map
is slightly closer to the star than the NNW peak---this is consistent with
the result obtained independently from the parametrized model fit.
The inferred displacement between the star and disk center 
results from the need to preferentially heat the SSE portion of the ring in 
order to reproduce the observed 30\% intensity enhancement of the SSE ansa at 
70 $\mu$m (see Figure \ref{fig2} and the discussion in \citet{sta04}). 
Therefore, the estimation accuracy of the displacement ($\pm 1$ AU) 
is dependent 
on these intensity ratios rather than the absolute pointing of the telescope.
Since the tau map embodies more observational data than the model fit,
we can refine our estimates of the other ring parameters; in particular
we obtain 180 AU for $r_{\rm out}$.

We have deprojected the tau map to produce a pole-on view of the disk, but
due to the
large inclination, the resulting errors were a factor of
three higher than for the sky projection.  The dynamic range was 
consequently insufficient for any further elucidation of the density 
structure and so we have not included the result here.

\section{Discussion}

Our 350 $\mu$m image is qualitatively different from the 450 $\mu$m
image published by \citet{hol03} in that it has the morphology of a symmetrical
ring as opposed to the horseshoe shape at 450 $\mu$m (ansae 
connected by a single arc). The latter was interpreted in terms of
a dust ``clump" suggestive of
material trapped in the mean motion resonance
of a large planet.  Since our 350 $\mu$m image is of similar signal to
noise ratio and resolution, we should have detected such a feature 
(at the $4\sigma$ level in this case).  More detailed examination shows that 
the arclike bridge at 450 $\mu$m has about the same relative
intensity with respect to the peak (0.5) as that of either arm of the
ring at 350 $\mu$m, whereas the western arm at 450 $\mu$m is
substantially weaker ($0.2\times{\rm peak}$). 
Thus rather than the eastern arm at 450 $\mu$m being an enhancement, the
western arm is actually a depression.  This argues against the presence
of a dust clump in the former location.

Our estimated disk parameters agree essentially with previous values
\citep{den00,hol03}, including the
opacity index ($\beta\simeq1$) which is consistent with relatively large
($\sim100$ $\mu$m) grains.  One possible discrepancy is that our opening
angle of $17^\circ$ implies a smaller vertical disk thickness
($\sim40$ AU) than previous estimates ($\sim80$ and 120 AU 
at 450 and 850 $\mu$m, respectively---\citet{hol03,den00}).
However, the estimated value is very sensitive to
PSF width, suggesting that at least some of the differences 
may be resolution-related.

Our optical depth mapping indicated an
$\sim8$ AU displacement of the ring center from the star. Such a displacement
was suggested by \citet{wya99} as a consequence of gravitational
perturbation by orbiting bodies, observable by the resulting pericenter
glow; the latter was invoked by \citet{sta04} to explain the
brightness asymmetry of the ansae.  Our estimated displacement 
corresponds to a forced eccentricity of approximately 0.06,
consistent with the findings of Stapelfeldt et al.  

In this scenario, one would expect an excess of material at apocenter
due to the slower orbital motion there. In order to see whether our data
show such an effect, we have simulated the tau map using
the results of dynamical modeling for a forced eccentricity of 0.06,
and the result is shown in the right hand panel of Figure \ref{fig3}.
Although the model tau map shows two equal peaks 
(in agreement with the estimated tau map), the NNW ansa (apocenter) is 
more radially extended than the SSW ansa since the orbits of individual 
grains are more dispersed there, and hence the integrated column density
at apocenter is greater.  A comparison of integrated column density in
the two halves of the ring (apocenter and pericenter), as inferred from the
model and estimated tau maps (truncated at a dust temperature of 35 K), 
indicates a predicted 12\% apocentric excess from the model, as compared to 
a value of $(14\pm3)$\% estimated from the observations.  The agreement
between these values supports the interpretation in terms of forced 
eccentricity.

If the orbit of the perturbing body (presumably a planet) is within
the inner boundary of the ring, 
our models indicate that a forced eccentricity of 0.06
would require an orbital eccentricity of
$\epsilon\simeq6/a_{\rm [AU]}$ for the planet, over a wide range of 
the planet's semi-major axis, $a$. For example, if the
inner edge of the ring is associated with the planet's 2:3 mean motion
resonance (analagous to Neptune and the Kuiper Belt in our own Solar
System), then $a\sim86$ AU and 
$\epsilon\sim0.07$.  A constraint on the planet's mass can be
derived from the timescale required to produce the observed perturbation;
the estimated age of Fomalhaut ($\sim200$ Myr) then implies a value of
$>1$ Earth mass.

\begin{figure}
\epsscale{.80}
\plotone{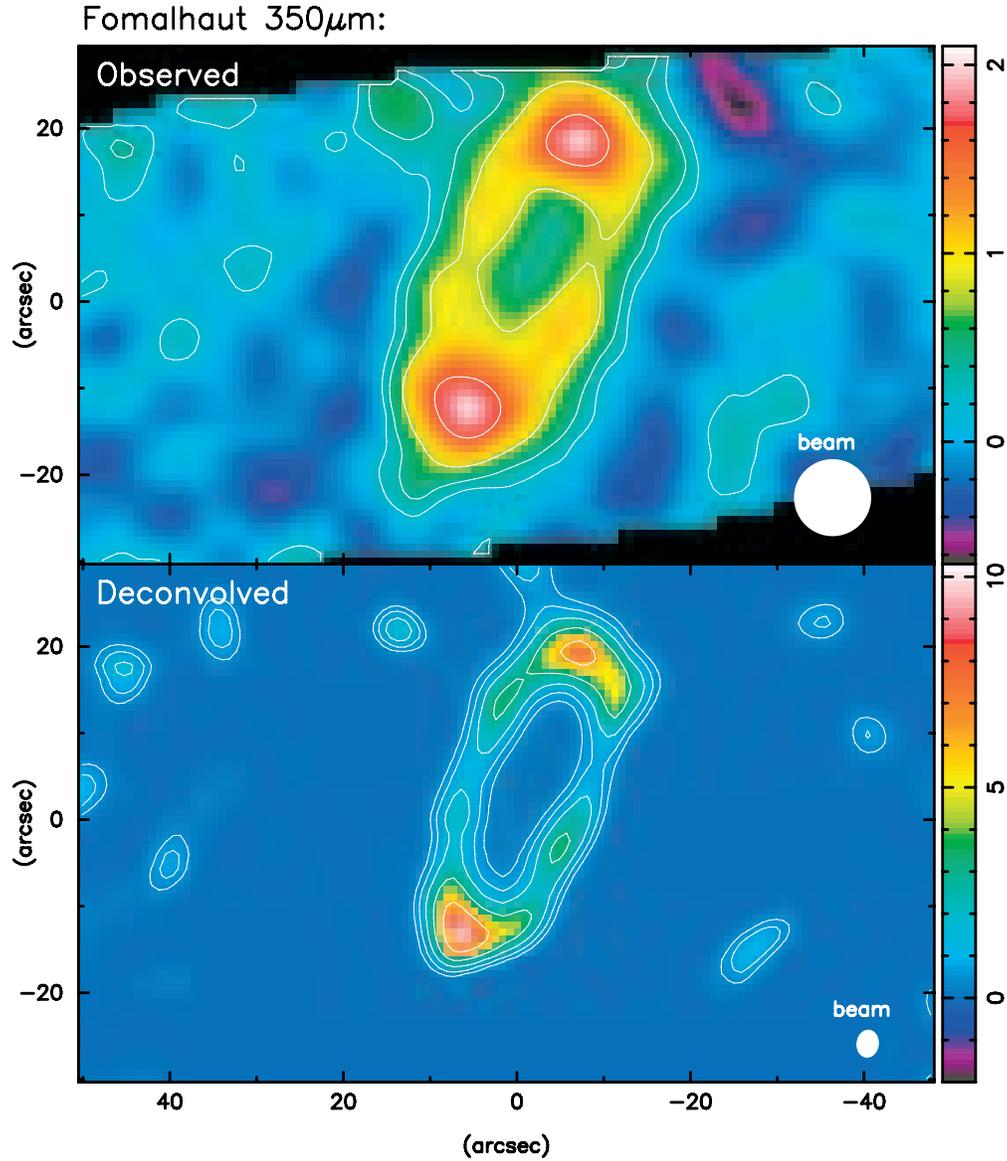}
\caption{Fomalhaut at 350 $\mu$m.  The upper and lower panels represent
the observed and deconvolved images, respectively.  In the observed image
the contours are at 0.2, 0.4, 0.8, \& 1.6 mJy ${\rm arcsec}^{-2}$ and the
RMS noise is 0.15 mJy ${\rm arcsec}^{-2}$. In the
deconvolved image the contours are at 0.4, 0.8, 1.6, 3.2, \& 6.4 
mJy ${\rm arcsec}^{-2}$ and the RMS noise is 0.18 mJy ${\rm arcsec}^{-2}$.
\label{fig1}}
\end{figure}

\begin{figure}
\plotone{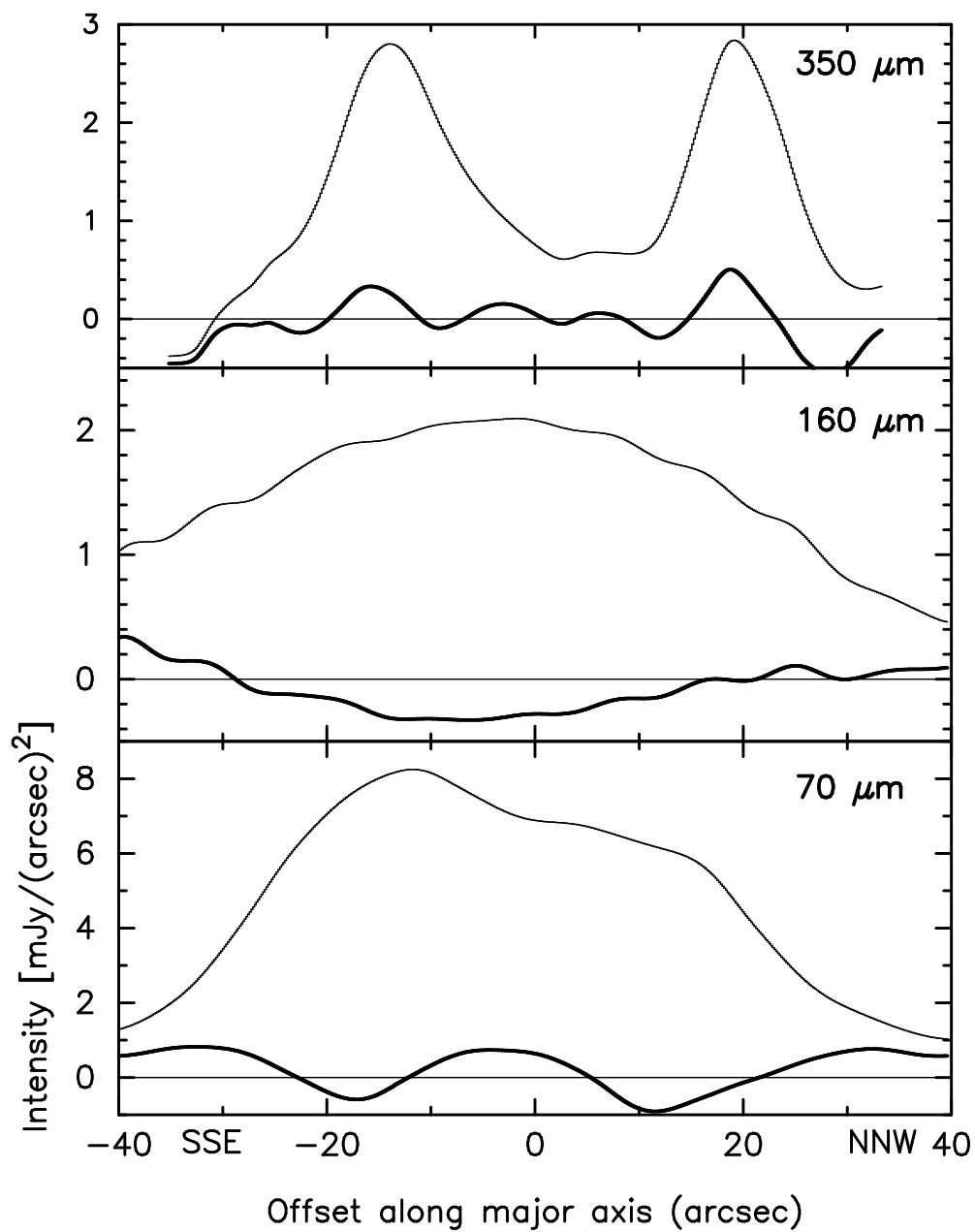}
\caption{The residuals of the uniform-ring model  
at 350, 160, and 70 $\mu$m (shown by the thick lines), as compared
to the observed images (thin lines).  The plots represent a slice through
the images along the major axis of the tilted ring.  The corresponding PSF 
widths (FWHM) were $9''$, $35''$, and $16''$ at 350, 160, and 70 $\mu$m,
respectively, and the absolute positional uncertainties were
$2''$, $3''$, and $3''$, respectively.
\label{fig2}}
\end{figure}

\begin{figure}
\plotone{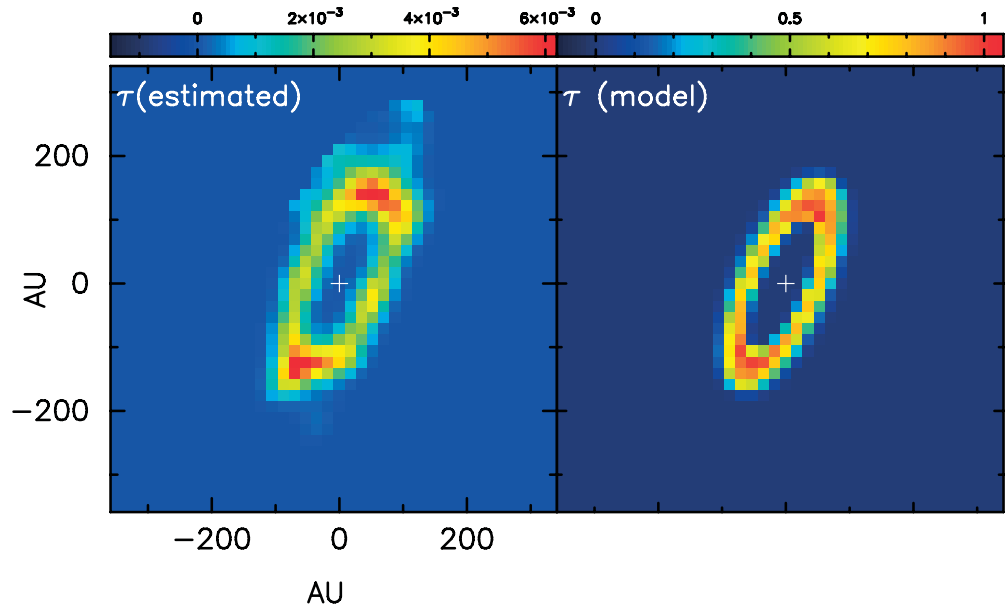}
\caption{Estimated and model tau maps of Fomalhaut,
projected onto the plane of the 
sky. {\em Left:\/} Estimated tau map, expressed at a reference wavelength of 
24 $\mu$m; the peak line-of-sight optical depth is 
$(6.61\pm0.26)\times10^{-3}$, with
the uncertainty increasing by a factor of 2 at the low intensity edges
of the ring.
{\em Right:\/} Theoretical tau map in relative units, obtained using a 
dynamical model based on a forced eccentricity of 0.06. This model map has
not been smoothed with an instrumental response, and therefore has
a sharper appearance than the estimated tau map at the left. \label{fig3}}
\end{figure}

An additional asymmetry is apparent in the estimated tau map of Figure
\ref{fig3}, in that the ring has a slight ``teardrop" shape, being more
pointed at pericenter.  This effect is apparent in the 350 $\mu$m 
data alone (see Figure \ref{fig1}) and also in a tau map made 
independently using only
the 24, 70, \& 160 $\mu$m {\em Spitzer\/} images (not reproduced here).
A possible explanation is warping of the disk, whereby the SSE portion of
the ring has greater inclination to the line of sight than the NNW portion.
To maintain such a warp would require at least two planets, since perturbations
by a single planet would result simply in the disk adjusting to that
planet's orbital plane.

Stronger constraints on the properties of the orbiting body (or bodies) could
be obtained from a more detailed knowledge of the azimuthal structure
of column density around the ring, which would require a more accurate
pole-on view than we were able to obtain with these data.  Such information
could, in principle, be obtained by supplementing the present data using
additional observations including, for example, the SED mode of MIPS, and
submillimeter mapping at higher signal to noise ratio.

\acknowledgments

We thank Dr. Matt Bradford for his assistance during the observations.
This work was performed by the Jet Propulsion Laboratory, 
California Institute of Technology, under contract with the 
National Aeronautics and Space Administration.  Research at the Caltech
Submillimeter Observatory is supported by NSF grant AST-0229008.

\clearpage

\begin{deluxetable}{ccl}
\tabletypesize{\scriptsize}
\tablecaption{Results of parameter fit.\label{table1}}
\tablewidth{0pt}
\tablehead{
\colhead{Parameter} & \colhead{Estimated value} & \colhead{Comment}
}
\startdata
$i$ & $68^\circ\pm 1^\circ$ & Inclination with respect to the plane of the
sky \\
$\theta$ & $-23^\circ\pm 1^\circ$ & Position angle of tilt axis \\
$r_{\rm in}$ & $113.0\pm 0.3$ AU & Inner radius \\
$r_{\rm out}$ & $238\pm 1$ AU & Outer radius \\
$\psi_{100}$ & $17^\circ\pm1^\circ$ & Opening angle for the flared-disk geometry \\
$\gamma$ & $0.10\pm0.05$ & Power-law index for radial variation of $\psi$ \\
$\tau(r_{\rm in})$ & $(1.6\pm0.1)\times 10^{-3}$ & Optical depth (normal to disk plane) at $r=r_{\rm in}$; $\lambda=24\,\mu$m \\
$\alpha$ & $1.73\pm0.01$ & Power-law index for radial variation of column density \\
$\beta$ & $1.02\pm0.01$ & Power-law index for wavelength variation of opacity \\
$\Delta x,\Delta y$ & $0.4\pm0.4;\,7.6\pm0.4$ & Offsets of dust ring from
star in AU (see text) \\
\enddata
\end{deluxetable}

\clearpage

\clearpage

\clearpage

\end{document}